\documentclass{article}
\usepackage{amssymb}
\renewcommand{\d}{\mathrm{d}}
\newcommand{\ud}{\mathrm{d}} 

\title{Quantization of N=1 and N=2 SUSY KdV models}
\author{Petr P. Kulish$^1$, Anton M. Zeitlin$^2$\\
St. Petersburg Department of Steklov Mathematical\\ 
Institute, Fontanka, 27, 
St. Petersburg, 191023, Russia\\
$^1$ kulish@pdmi.ras.ru $^2$ zam@math.ipme.ru\\
$^2$ http://www.ipme.ru/zam.html}
\begin{document}
\maketitle

\begin{abstract}
The quantization procedure for both N=1 and N=2 supersymmetric 
Korte\-weg-de Vries 
(SUSY KdV) hierarchies is constructed. Namely, the quantum counterparts of 
the monodromy matrices, built by means of the integrated vertex operators, 
are shown to satisfy a specialization of reflection equation, leading 
to the quantum integrable theory. The relation of such models to the study 
of integrable perturbed superconformal and topological models is 
discussed.       
\end{abstract}

\section{Introduction}
\hspace*{5mm}The Superconformal Field Theory (SCFT) is very important in many areas of Mathematical Physics, 
such as Superstring theory, where it appears as a basic part of the String Worldsheet Physics 
or the theory of two-dimensional 
solvable lattice models, e.g. tricritical Ising model as the well known example. 
The symmetries of SCFT are described by the generators of the Superconformal Algebra (SCA). This algebra 
is a supersymmetric extension of the infinite dimensional Virasoro algebra, which in its turn is 
a central extension of the algebra of vector fields on a circle.\\ 
\hspace*{5mm}There are two types of SCA which are especially important in physical problems, 
these are the so-called N=1 and N=2 SCA, named by the number 
of fermionic degrees of freedom in the string worldsheet superspace. 
The infinite dimensionality of this symmetry allows to look for integrable structures 
inside SCFT, that is infinite families of pairwise commuting integrals of motion (IM). 
Motivated by the ``physical'' reasons we impose also the condition of the supersymmetric (SUSY) 
invariance of these families, i.e. the SUSY generator(s) should be included in these families.\\   
\hspace*{5mm}In this paper we review some 
results concerning the construction of these families for both N=1 and N=2 SCFT. 
The key idea is that on the classical level the integrable structures are provided by the so-called 
second Hamiltonian structure of N=1 and N=2 SUSY KdV correspondingly. In order to build infinite 
family of pairwise commuting quantum IM, we need to quantize the corresponding integrable models.\\
\hspace*{5mm}Here we review the quantization procedure for both N=1 and N=2 SUSY KdV according to the 
results from the papers \cite{susykdv}, \cite{toda-kdv}, \cite{tmph}. 
In the first part (Sec. 2) we consider the classical theory of SUSY N=1
KdV system, based on the twisted affine superalgebra $C(2)^{(2)}\simeq 
sl(1|2)^{(2)}\simeq osp(2|2)^{(2)}$. We introduce the supersymmetric
Miura transformation and monodromy matrix associated with the 
corresponding L-operator.
Then the auxiliary $\mathbf{L}$-matrices are constructed, 
which satisfy the quadratic Poisson bracket relation. 
In Sec. 4 we will show that quantum counterparts of these matrices coincide with a 
vertex-operator-represented quantum R-matrix.
The quantum version of the Miura transformation, i.e. 
the free field representation of the superconformal algebra, is given in 
Sec. 3. In Sec. 4 the quantum $C_{q}(2)^{(2)}$ superalgebra \cite{tolstkhor} 
is introduced.
Then it is shown that the corresponding quantum R-matrix can be represented
by two vertex operators, satisfying Serre relations of lower Borel subalgebra
of $C_{q}(2)^{(2)}$ and in the classical limit it coincides with 
$\mathbf{L}$-matrix.
The vertex-operator-represented quantum R-matrix $\mathbf{L}^{(q)}$
satisfies the so-called RTT-relation, which gives a 
possibility to consider the 
model from a point of view of Quantum Inverse Scattering Method (QISM) 
\cite{leshouches}, \cite{kulsklyan}.\\
\hspace*{5mm}
In the next four sections we switch to the study of the SUSY N=2 KdV quantization procedure.
The N=2 KdV model is based on the $sl^{(1)}(2|1)$ affine  superalgebra. 
The classical version of the associated monodromy matrix is represented by means of familiar 
P-exponential but the exponent consists not only of generators of simple roots and the squares of fermionic 
ones with appropriate exponential multipliers, but also generators, corresponding to more complicated 
composite roots (see Sec. 5). 
We prove that (as it was in the N=1 case) all these terms, corresponding to the composite roots 
disappear from the first iteration of the quantum generalization of the P-exponential; we also show that 
this ``quantum'' P-exponential coincide with the reduced universal R-matrix of 
$\hat{sl}_q(2|1)$ quantum affine superalgebra (see Sec. 6,7).
We demonstrate that the supertraces of the quantum version of the monodromy matrix in different evaluation 
representations, the so-called transfer-matrices (by the obvious analogy with the lattice case) 
commute with the 
supersymmetry generators. Hence, these generators can be included in the family of the integrals of motion 
both on classical and the quantum level (see Sec. 8).\\ 
\hspace*{5mm}In the last section we discuss the relation of N=1 and N=2 SUSY KdV models 
to the superconformal/topological theories and their integrable perturbations.

\section{Integrable SUSY N=1 KdV hierarchy}
\hspace*{5mm}The SUSY N=1 KdV system can be constructed by means of the Drinfeld-Sokolov
reduction applied to the $C(2)^{(2)}$ twisted affine superalgebra 
\cite{inami}. 
The corresponding
$\mathcal{L}$-operator has the following form:
\begin{eqnarray}
\tilde{\mathcal{L}}_F=D_{u,\theta} 
-D_{u,\theta}\Phi (h_1+h_2)-\lambda({e_1}^{+}+{e_2}^{+}+{e_1}^{-}-{e_2}^{-}),
\end{eqnarray}
where $D_{u,\theta} =\partial_\theta + \theta \partial_u$ 
is a superderivative, the variable 
$u$ lies on a cylinder of circumference $2\pi$, $\theta$ 
is  a  Grassmann  variable, $\Phi(u,\theta)=\phi(u) - 
\frac{i} {\sqrt{2}}\theta\xi(u)$ 
is a bosonic superfield;
$h_1$, $h_2$, ${e_1}^{\pm}$, ${e_2}^{\pm}$ are the Chevalley 
generators of C(2) with the following commutation relations:
\begin{eqnarray} 
&&[h_1,h_2]=0,\quad [h_1,{e_2}^{\pm}]=\pm{e_2}^{\pm},
\quad [h_2,{e_1}^{\pm}]=\pm{e_1}^{\pm},\\
&&ad^{2}_{{e_1}^{\pm}}{e_2}^{\pm}=0, \quad ad^{2}_{{e_2}^{\pm}}{e_1}^{\pm}=0, 
\nonumber\\
&&[h_{\alpha},{e_{\alpha}}^{\pm}]=0 \quad(\alpha =1,2),\quad
[{e_{\beta}}^{\pm},{e_{\beta'}}^{\mp}]=\delta_{\beta, \beta'}h_{\beta}\quad
(\beta , \beta' =1,2),\nonumber
\end{eqnarray}
where the supercommutator [,] is defined as follows:
$[a,b]\equiv ad_{a} b\equiv ab- (-1)^{p(a)p(b)}ba$, where parity $p$
is equal to 1 for odd elements and is equal to 0 for even ones.
In the particular case of C(2) $h_{1,2}$ are even and $e^{\pm}_{1,2}$
are odd. 
The operator (1) can be considered as more general one, taken in the 
evaluation representation of $C(2)^{(2)}$: 
\begin{eqnarray}
\mathcal{L}_F=D_{u,\theta} 
-D_{u,\theta}\Phi h_{\alpha}-(e_{{\delta-\alpha}}+ e_{{\alpha}}),
\end{eqnarray}
where $h_{\alpha}$, $e_{{\delta-\alpha}}$, $e_{{\alpha}}$ are the 
Chevalley generators of $C(2)^{(2)}$ with such commutation 
relations: 
\begin{eqnarray}
&&[h_{\alpha_1},h_{\alpha_0}]=0,\quad [h_{\alpha_0},e_{\pm\alpha_1}]= 
\mp e_{\pm\alpha_1},\quad [h_{\alpha_1},e_{\pm\alpha_0}]= 
\mp e_{\pm\alpha_0},\\
&&[h_{\alpha_i},e_{\pm\alpha_i}]=\pm e_{\pm\alpha_i},\quad
[e_{\pm\alpha_i}, e_{\mp\alpha_j}]=\delta_{i,j}h_{\alpha_i},\quad (i,j=0,1),
\nonumber\\
&&ad^{3}_{e_{\pm\alpha_0}} e_{\pm\alpha_1}=0, \quad
ad^{3}_{e_{\pm\alpha_1}} e_{\pm\alpha_0}=0\nonumber
\end{eqnarray}
where $p(h_{\alpha_{0,1}})=0$, $p(e_{\pm \alpha_{0,1}})=1$ and 
$\alpha_1\equiv \alpha$, $\alpha_0\equiv\delta-\alpha$.
The Poisson brackets for the field $\Phi$, obtained by means of the 
Drinfeld-Sokolov
reduction are:
\begin{equation}
\{D_{u,\theta}\Phi(u,\theta), D_{u',\theta'}\Phi(u',\theta')\}=  
D_{u,\theta}(\delta(u-u')(\theta-\theta'))
\end{equation}
and the following boundary conditions are imposed on the components of 
$\Phi$: $\phi(u+2\pi)=\phi(u)+2\pi i p$, $\xi(u+2\pi)=\pm\xi(u)$.
The $\mathcal{L}_F$-operator is written in the Miura form, making a gauge
transformation one can obtain a new superfield $\mathcal{U}(u,\theta)\equiv
D_{u,\theta}\Phi(u,\theta)\partial_u\Phi(u,\theta)-D_{u,\theta}^3
\Phi(u,\theta)=-\theta U(u)-i\alpha(u)/\sqrt{2}$, 
where $U$ and $\alpha$ generate the superconformal algebra under the Poisson 
brackets:
\begin{eqnarray}
\{U(u),U(v)\}&=&
 \delta'''(u-v)+2U'(u)\delta(u-v)+4U(u)\delta'(u-v),\\
\{U(u),\alpha(v)\}&=&
 3\alpha(u)\delta'(u-v) + \alpha'(u)\delta(u-v),\nonumber\\
\{\alpha(u),\alpha(v)\}&=&
 2\delta''(u-v)+2U(u)\delta(u-v)\nonumber.
\end{eqnarray} 
Using one of the corresponding infinite family of IM, which are in involution
under the Poisson brackets 
(these IM could be extracted from the monodromy matrix 
of $\mathcal{L}_B$-operator, see below)\cite{gervais}:
\begin{eqnarray}
I^{(cl)}_1&=&\frac{1}{2\pi}\int U(u)\d u,\\
I^{(cl)}_3&=&\frac{1}{2\pi}\int
\Big(U^2(u)+\alpha(u)\alpha'(u)/2\Big)\d u,\nonumber\\
I^{(cl)}_5&=&\frac{1}{2\pi}\int
\Big(U^3(u)-(U')^2(u)/2-\alpha'(u)\alpha''(u)/4-
\alpha'(u)\alpha(u)U(u)\Big)\d u,\nonumber\\
& &   .\qquad.\qquad.\nonumber
\end{eqnarray}
one can obtain an evolution equation; for example, taking $I_2$
we get the SUSY N=1 KdV equation \cite{mathieu}:
$\mathcal{U}_t=-\mathcal{U}_{uuu}+3(\mathcal{U} D_{u,\theta}\mathcal{U})_u$ 
and in components:
$
U_t=-U_{uuu}-6UU_u - \frac{3}{2}\alpha\alpha_{uu}$, 
$\alpha_t=-4\alpha_{uuu}-3(U\alpha)_u.$
As we have noted in the introduction one can show that the 
IM are invariant under supersymmetry transformation generated by
$\int_0^{2\pi}\d u\alpha(u)$.\\
\hspace*{5mm}In order to construct the so-called monodromy 
matrix we introduce the 
$\mathcal{L}_B$-operator, equivalent to the $\mathcal{L}_F$ one:
\begin{eqnarray}
\mathcal{L}_B=\partial_u-\phi'(u)h_{\alpha_1}
+(e_{\alpha_1}+e_{\alpha_0}-\frac{i}{\sqrt{2}}\xi h_{\alpha_1})^2
\end{eqnarray}
The equivalence can be easily established if one considers the linear 
problem associated with the $\mathcal{L}_F$-operator: $\mathcal{L}_F\chi(u,
\theta)=0$ (we consider this operator acting in some representation of
$C(2)^{(2)}$ and $\chi(u,
\theta)$ is the vector in this representation). 
Then, expressing $\chi(u,\theta)$ in components:  
$\chi(u,\theta)$=$\chi_0(u)+\theta\chi_1(u)$, we find:
$\mathcal{L}_B\chi_0=0$ and $\chi_1=(e_{\alpha_1}+e_{\alpha_0}-\frac{i}{\sqrt{2}}\xi h_{\alpha_1})\chi_0$.\\
\hspace*{5mm} The solution to the equation $\mathcal{L}_B\chi_0=0$ can be written in the following way:
\begin{eqnarray}
\chi_0(u)&=&
e^{\phi(u)h_{\alpha_1}}P\exp\int_0^u \d u'\Big(\frac{i}{\sqrt{2}}
\xi(u')e^{-\phi(u')}e_{\alpha_1}\\
&-&\frac{i}{\sqrt{2}}
\xi(u')e^{\phi(u')}e_{\alpha_0}
-e^2_{\alpha_1}e^{-2\phi(u')}-
e^2_{\alpha_0}e^{2\phi(u')}-[e_{\alpha_1},e_{\alpha_0}]
\Big)\eta,\nonumber
\end{eqnarray}
where $\eta$ is a constant vector in the corresponding representation 
of $C(2)^{(2)}$.
Therefore we can define the monodromy matrix in the following way:
\begin{eqnarray}
\mathbf{M}&=&e^{2\pi i ph_{\alpha_1}}
P\exp\int_0^{2\pi} \d u\Big(\frac{i}{\sqrt{2}}
\xi(u)e^{-\phi(u)}e_{\alpha_1}\\
&-&\frac{i}{\sqrt{2}}
\xi(u)e^{\phi(u)}e_{\alpha_0}
-e^2_{\alpha_1}e^{-2\phi(u)}-
e^2_{\alpha_0}e^{2\phi(u)}-[e_{\alpha_1},e_{\alpha_0}]
\Big).\nonumber
\end{eqnarray}
Introducing then (as in \cite{1}) the auxiliary $\mathbf{L}$-operators:
$\mathbf{L}=e^{-\pi ip h_{\alpha_1}}\mathbf{M}$ we find that in the 
evaluation representation
(when $\lambda$, the spectral parameter appears) the following Poisson 
bracket relation is satisfied \cite{solitons}:
\begin{eqnarray}\label{eq:LLr}
\{\mathbf{L}(\lambda)\otimes_{,}\mathbf{L}(\mu)\}=
[\mathbf{r}(\lambda\mu^{-1}),\mathbf{L}(\lambda)\otimes \mathbf{L}(\mu)],
\end{eqnarray}
where $\mathbf{r}(\lambda\mu^{-1})$ is trigonometric $C(2)^{(2)}$ r-matrix 
\cite{shadr}.
From this relation one obtains that the 
supertraces of monodromy matrices 
$\mathbf{t}(\lambda)=str\mathbf{M}(\lambda)$ 
commute under the Poisson bracket:
$\{\mathbf{t}(\lambda),\mathbf{t}(\mu)\}=0$.
Expanding $\log(\mathbf{t}(\lambda))$ in 
$\lambda$ in the evaluation representation corresponding
to the defining 3-dimensional 
representation of $C(2)$ $\pi_{1/2}$ we find:
\begin{eqnarray}
\log(\mathbf{t}_{1/2}(\lambda))=-\sum^{\infty}_{n=1}c_n 
I^{(cl)}_{2n-1}{\lambda}^{-4n+2},\qquad \lambda\to\infty 
\end{eqnarray}
where $c_1=\frac{1}{2}$, $c_n=\frac{(2n-3)!!}{2^n n!}$ for $n>1$. 
So, one can obtain the IM from the
supertrace of the monodromy matrix. Using the Poisson bracket relation
for these supertraces with different values of spectral parameter 
(see above) we find that infinite family of IM is involutive, as it was 
mentioned earlier.
\section{Free field representation of superconformal algebra}
\hspace*{5mm}In this section we begin to build quantum counterparts of the introduced classical objects. 
We will start from the quantum Miura transformation, the free 
field representation of the superconformal algebra \cite{SCFT}:
\begin{eqnarray}
-\beta^2T(u)&=&:\phi'^2(u):-(1-\beta^2/2)\phi''(u)+\frac{1}{2}:\xi\xi'(u):+\frac{\epsilon\beta^2}{16}\\ 
\frac{i^{1/2}\beta^2}{\sqrt{2}}G(u)&=&\phi '\xi(u)-(1-\beta^2/2)\xi '(u),
\nonumber
\end{eqnarray}
where
\begin{eqnarray}
&&\phi(u)=iQ+iPu+\sum_n\frac{a_{-n}}{n}e^{inu},\qquad
\xi(u)=i^{-1/2}\sum_n\xi_ne^{-inu},\\
&&[Q,P]=\frac{i}{2}\beta^2 ,\quad 
[a_n,a_m]=\frac{\beta^2}{2}n\delta_{n+m,0},\qquad
\{\xi_n,\xi_m\}=\beta^2\delta_{n+m,0}.\nonumber
\end{eqnarray}
Recall that there are two types of boundary conditions on 
$\xi$: $\xi(u+2\pi)=\pm\xi(u)$. The sign ``+'' corresponds 
to the R sector,the case
when $\xi$ is integer modded, the ``--'' sign corresponds to the NS sector and
$\xi$ is half-integer modded. The variable $\epsilon$ in (13)
is equal to zero
in the R case and equal to 1 in the NS case.\\
One can expand $T(u)$ and $G(u)$ by modes in such a way: $
T(u)=\sum_nL_{-n}e^{inu}-\frac{\hat{c}}{16}$, $G(u)=\sum_nG_{-n}e^{inu}
$,
where  $\hat{c}=5-2(\frac{\beta^2}{2}+\frac{2}{\beta^2})$  and $L_n,G_m$ 
generate the superconformal algebra:
\begin{eqnarray}
[L_n,L_m]&=&(n-m)L_{n+m}+\frac{\hat{c}}{8}(n^3-n)\delta_{n,-m}, \quad
\lbrack L_n,G_m\rbrack=(\frac{n}{2}-m)G_{m+n}\nonumber\\
\lbrack G_n,G_m\rbrack&=&2L_{n+m}+\delta_{n,-m}\frac{\hat{c}}{2}(n^2-1/4).
\end{eqnarray}
In the classical limit  $c\to -\infty$ (the same is $\beta^2\to 0$) 
the following substitution:
$
T(u)\to-\frac{\hat{c}}{4}U(u)$, 
$G(u)\to-\frac{\hat{c}}{2\sqrt{2i}}\alpha(u)$,
$[,]\to \frac{4\pi}{i\hat{c}}\{,\}
$
reduce the above algebra to the Poisson bracket algebra of 
SUSY N=1 KdV theory.\\ 
\hspace*{5mm}Let now $F_p$ be the highest weight 
module over the oscillator algebra of 
$a_n$, $\xi_m$ with the highest weight vector (ground state) $|p\rangle$ 
determined by the 
eigenvalue of $P$ and nilpotency condition of the action of the positive modes:
$
P|p\rangle=p|p\rangle,\quad 
a_n|p\rangle=0, \quad \xi_m|p\rangle=0$ where $n,m > 0$.
In the case of the R sector the highest weight becomes doubly degenerate
due to the presence of zero mode $\xi_0$. So, there are two ground states
$|p,+\rangle$ and $|p,-\rangle$: $|p,+\rangle = \xi_0|p,-\rangle$.
Using the above free field representation of the superconformal algebra
one can obtain that for generic $\hat{c}$ and $p$, $F_p$ is isomorphic to the 
super-Virasoro module with the highest weight vector $|p\rangle$:
$
L_0|p\rangle=\Delta_{NS}|p\rangle,$ where $\Delta_{NS}=
(p/\beta)^2 + (\hat{c}-1)/16$ in the NS sector and module with two highest weight vectors in the Ramond case:
$
L_0|p,\pm\rangle=\Delta_{R}|p,\pm\rangle,\quad\Delta_{R}=
(p/\beta)^2 + \hat{c}/16,\quad
|p,+\rangle=(\beta^2/\sqrt{2}p)G_0|p,-\rangle.
$
The space $F_p$, now considered as super-Virasoro module, splits 
into the sum of finite-dimensional subspaces, determined by the 
value of $L_0$: $
F_p=\oplus^{\infty}_{k=0}F_p^{(k)}$, $L_0 F_p^{(k)}=(\Delta + k) F_p^{(k)}$.
The quantum versions of local integrals of motion should act invariantly on 
the subspaces $F_p^{(k)}$. Thus, the diagonalization
of IM reduces (in a given subspace $ F_p^{(k)}$) to the finite purely 
algebraic problem,
which however rapidly become rather complicated for large $k$. It should 
be noted also that in the case of the 
Ramond sector supersymmetry generator $G_0$ commute with 
IM, so IM act in $|p,+\rangle$ and $|p,-\rangle$ independently, 
without mixing of these
two ground states (unlike the super-KdV case \cite{super-kdv}).
\section {Quantum monodromy matrix and RTT-relation}
\hspace*{5mm}In this part of the work we will consider the quantum $C_q(2)^{(2)}$ 
R-matrix and show that the vertex operator representation of the lower Borel 
subalgebra of $C_q(2)^{(2)}$ allows to represent this R-matrix in the
P-exponent like form which in the classical limit coincide with the
auxiliary L-operator.
$C_q(2)^{(2)}$ is a quantum superalgebra with the following commutation 
relations \cite{tolstkhor}:
\begin{eqnarray}
&&[h_{\alpha_0},h_{\alpha_1}]=0,\quad [h_{\alpha_0},e_{\pm\alpha_1}]= 
\mp e_{\pm\alpha_1},\quad [h_{\alpha_1},e_{\pm\alpha_0}]= 
\mp e_{\pm\alpha_0},\\
&&[h_{\alpha_i},e_{\pm\alpha_i}]=\pm e_{\pm\alpha_i}\quad (i=0,1),\quad
[e_{\pm\alpha_i}, e_{\mp\alpha_j}]=\delta_{i,j}[h_{\alpha_i}]\quad (i,j=0,1),
\nonumber\\
&&[e_{\pm\alpha_1},[e_{\pm\alpha_1},[e_{\pm\alpha_1},
e_{\pm\alpha_0}]_{q}]_{q}]_{q}=0,\quad 
[[[e_{\pm\alpha_1},e_{\pm\alpha_0}]_{q},e_{\pm\alpha_0}]_{q},
e_{\pm\alpha_0}]_{q}=0,
\nonumber
\end{eqnarray}
where $[x]=\frac{q^x-q^{-x}}{q-q^{-1}}$, $p(h_{\alpha_{0,1}})=0$, 
$p(e_{\pm \alpha_{0,1}})=1$ and 
q-supercommutator is defined in the following way:
$[e_{\gamma},e_{\gamma'}]_{q}\equiv e_{\gamma}e_{\gamma'} - 
(-1)^{p(e_{\gamma})p(e_{\gamma'})}
q^{(\gamma,\gamma')}e_{\gamma'}e_{\gamma}$, $q=e^{i\pi\frac{\beta^2}{2}}$.
The corresponding coproducts are:
\begin{eqnarray}
&&\Delta(h_{\alpha_j})=h_{\alpha_j}\otimes 1 + 1\otimes h_{\alpha_j},
\quad
\Delta(e_{\alpha_j})=e_{\alpha_j}\otimes q^{h_{\alpha_j}}+1\otimes 
e_{\alpha_j},\\
&&\Delta(e_{-\alpha_j})=e_{-\alpha_j}
\otimes 1+q^{-h_{\alpha_j}} \otimes e_{-\alpha_j}.\nonumber
\end{eqnarray}
The associated R-matrix can be expressed in such a way \cite{tolstkhor}:
$R=KR_{+}R_{0}R_{-}$, where $K=q^{h_{\alpha}\otimes h_{\alpha}}$, 
$R_{+}=\prod_{n\ge 0}^{\to}R_{n\delta+\alpha}$, 
$R_{-}=\prod_{n\ge 1}^{\gets}R_{n\delta-\alpha}$,
$R_{0}=\exp((q-q^{-1})\sum_{n>0}d(n)e_{n\delta}\otimes e_{-n\delta}).$
Here $R_{\gamma}=\exp_{(-q^{-1})}(A(\gamma)(q-q^{-1})(e_{\gamma}
\otimes e_{-\gamma}))$ and coefficients $A$ and $d$ are defined as follows:
$A(\gamma)=\{(-1)^n$ if $\gamma=n\delta+\alpha; (-1)^{n-1}$ if 
$\gamma=n\delta-\alpha\}$, $d(n)=\frac{n(q-q^{-1})}{q^n-q^{-n}}$.
The generators $e_{n\delta}$, $e_{n\delta\pm\alpha}$ are defined via the 
q-commutators of Chevalley generators. In the following we will need  
the expressions only for simplest ones: 
$e_{\delta}=[e_{\alpha_0}, e_{\alpha_1}]_{q^{-1}}$ and 
$e_{-\delta}=[e_{-\alpha_1},e_{-\alpha_0}]_q$. The elements 
$e_{n\delta\pm\alpha}$ are expressed as multiple commutators of 
$e_{\delta}$ with corresponding Chevalley generators, $e_{n\delta}$ ones 
have more complicated form \cite{tolstkhor}.\\    
\hspace*{5mm}Let's introduce the reduced R-matrix $\bar{R}\equiv K^{-1}R$.
Using all previous information one can write $\bar{R}$ as
$\bar{R}(\bar{e}_{\alpha_i},\bar{e}_{-\alpha_i})$, where
$\bar{e}_{\alpha_i}=e_{\alpha_i}\otimes 1$ and
$\bar{e}_{-\alpha_i}=1\otimes e_{-\alpha_i}$, because it is
represented as power series of these elements.
After this necessary background we will 
introduce vertex operators and using the fact that they
satisfy the Serre relations of the lower Borel subalgebra of $C_q(2)^{(2)}$  
we will prove that the reduced R-matrix, represented by the vertex operators 
has the properties of the P-exponent.
So, the vertex operators are:
\begin{eqnarray}
V_1=\frac{1}{q^{-1}-q}\int \d\theta \int^{u_{1}}_{u_{2}} \d u :e^{-\Phi}:,
\quad 
V_0=\frac{1}{q^{-1}-q}\int \d\theta \int^{u_{1}}_{u_{2}} \d u :e^{\Phi}:,
\end{eqnarray}
where $2\pi \ge u_1 \ge\ u_2\ge 0$, $\Phi=\phi(u) - 
\frac{i} {\sqrt{2}}\theta\xi(u)$ is a superfield and normal ordering 
here means that
$:e^{\pm\phi(u)}:=
\exp\Big(\pm\sum_{n=1}^{\infty}\frac{a_{-n}}{n}e^{inu}\Big)
\exp\Big(\pm i(Q+Pu)\Big)\exp\Big(\mp\sum_{n=1}^{\infty}\frac{a_{n}}{n}e^{-inu}
\Big)$. 
One can show via the standard contour technique that these 
operators satisfy the
same commutation relations as $e_{\alpha_1}$, $e_{\alpha_0}$ correspondingly.\\
\hspace*{5mm}Then, following \cite{4} one can show 
(using the fundamental property of the universal R-matrix: 
$(I\otimes\Delta)R=R^{13}R^{12}$) 
that the reduced R-matrix has the following 
property: 
\begin{equation}
\bar{R}(\bar{e}_{\alpha_i},e'_{-\alpha_i}+e''_{-\alpha_i})=
\bar{R}(\bar{e}_{\alpha_i},e'_{-\alpha_i})\bar{R}(\bar{e}_{\alpha_i},
e''_{-\alpha_i}),
\end{equation} 
where $e'_{-\alpha_i}=1\otimes q^{-h_{\alpha_i}}\otimes 
e_{-\alpha_i}$, 
$e''_{-\alpha_i}=1\otimes e_{-\alpha_i}\otimes 1$,
$\bar{e}_{\alpha_i}=e_{\alpha_i} \otimes 1\otimes 1$. 
The commutation relations between them are:
\begin{eqnarray}
e'_{-\alpha_i}\bar{e}_{\alpha_j}&=&-\bar{e}_{\alpha_j}e'_{-\alpha_i},\quad
e''_{-\alpha_i}\bar{e}_{\alpha_j}=-\bar{e}_{\alpha_j}e''_{-\alpha_i},\\
e'_{-\alpha_i}e''_{-\alpha_j}&=&-q^{b_{ij}}e''_{-\alpha_j}e'_{-\alpha_i},
\nonumber
\end{eqnarray}
where $b_{ij}$ is the symmetric matrix with the following elements:
$b_{00}=b_{11}=-b_{01}=1$. Now, denoting $\mathbf{\bar{L}}^{(q)}(u_2, u_1)$ 
the reduced R-matrix with $e_{-\alpha_i}$ represented by $V_{i}$, we find,
using the above property of $\bar{R}$ with $e'_{-\alpha_i}$ replaced by 
appropriate vertex operators: 
$\mathbf{\bar{L}}^{(q)}(u_3, u_1)=\mathbf{\bar{L}}^{(q)}(u_2, u_1)
\mathbf{\bar{L}}^{(q)}(u_3, u_2)$ with $u_1 \ge u_2\ge u_3 $. 
So, $\mathbf{\bar{L}}^{(q)}$ has the property of P-exponent. 
But because of singularities in the operator 
products of vertex operators it can not be written in the usual P-ordered
form. Thus, we propose a new notion, the quantum P-exponent: 
\begin{equation}
\mathbf{\bar{L}}^{(q)}(u_1, u_2)=Pexp^{(q)}\int^{u_1}_{u_2}\d u \int\d \theta
(e_{\alpha_1} :e^{-\Phi}:+e_{\alpha_0}:e^{\Phi}:).
\end{equation}
Introducing new object: $\mathbf{L}^{(q)}\equiv e^{i\pi Ph_{\alpha_1}}
\mathbf{\bar{L}}^{(q)}(0,2\pi)$, which coincides with R-matrix with 
$1\otimes h_{\alpha_1}$ replaced by $2P/\beta^2$ and 
$1\otimes e_{-\alpha_1}$, $1\otimes e_{-\alpha_0}$ replaced by $V_1$ and 
$V_0$ (with integration from 0 to $2\pi$) correspondingly, 
we find that it satisfies the well-known RTT-relation:
\begin{eqnarray}
&&\mathbf{R}(\lambda\mu^{-1})
\Big(\mathbf{L}^{(q)}(\lambda)\otimes \mathbf{I}\Big)\Big(\mathbf{I}
\otimes \mathbf{L}^{(q)}(\mu)\Big)\\
&&=(\mathbf{I}\otimes \mathbf{L}^{(q)}(\mu)\Big)
\Big(\mathbf{L}^{(q)}(\lambda)\otimes \mathbf{I}\Big)\mathbf{R}
(\lambda\mu^{-1}), \nonumber
\end{eqnarray}
where the dependence on $\lambda,\mu$ means that we are considering
$\mathbf{L}^{(q)}$-operators in the evaluation representation of 
$C_{q}(2)^{(2)}$. 
The quantum counterpart of the monodromy matrix $\mathbf{M}^{(q)}=e^{\sum_{i=1,2}\pi i p_ih_{\alpha_i}}\bar{\mathbf{L}}^{(q)}$ satisfies the reflection equation \cite{reflection}:
\begin{eqnarray}
\mathbf{\tilde{R}}_{12}(\lambda\mu^{-1})
\mathbf{M}_1^{(q)}(\lambda)F^{-1}_{12}\mathbf{M}^{(q)}_2(\mu)=\mathbf{M}_2^{(q)}(\mu)F^{-1}_{12}
\mathbf{M}_1^{(q)}(\lambda)\mathbf{R}_{12}(\lambda\mu^{-1}),
\end{eqnarray}
where $F=K^{-1}$, the Cartan's factor from the universal R-matrix, 
$\mathbf{\tilde{R}}_{12}(\lambda\mu^{-1})=F^{-1}_{12}\mathbf{R}_{12}(\lambda\mu^{-1})F_{12}$ and labels 1,2 denote the position of multipliers in the tensor 
product.
Thus the supertraces of monodromy matrices, the 
transfer matrices
$\mathbf{t}^{(q)}(\lambda)\equiv str (\mathbf{M}^{(q)}(\lambda))$ 
commute
\begin{eqnarray}
[\mathbf{t}^{(q)}(\lambda),\mathbf{t}^{(q)}(\mu)]=0,
\end{eqnarray}
giving the quantum integrability.
\hspace*{5mm}Now we will show that in the classical
limit ($q\to1$) the $\mathbf{L}^{(q)}$-operator will give the 
auxiliary $\mathbf{L}$-matrix defined in the Sec. 2.  
We will use the P-exponent property of $\mathbf{\bar{L}}^{(q)}(0, 2\pi)$.
Let's decompose $\mathbf{\bar{L}}^{(q)}(0, 2\pi)$ in the following way:
$
\mathbf{\bar{L}}^{(q)}(0, 2\pi)=
\lim_{N\to\infty}\prod_{m=1}^{N}\mathbf{\bar{L}}^{(q)}(x_{m-1},x_{m})
$, where we divided  the interval $[0,2\pi]$ into infinitesimal 
intervals $[x_m,x_{m+1}]$
with $x_{m+1}-x_m=\epsilon=2\pi/N$.
Let's find the terms that can give contribution of the first order
in $\epsilon$ in $\mathbf{\bar{L}}^{(q)}(x_{m-1},x_{m})$. 
In this analysis we will need the operator product expansion of 
vertex operators:
\begin{eqnarray}
&&\xi(u)\xi(u')=
-\frac{i\beta^2}{(iu-iu')}+\sum_{k=1}^{\infty}c_k(u)(iu-iu')^k,\\
&&:e^{a\phi(u)}::e^{b\phi(u')}:=(iu-iu')^{\frac{ab\beta^2}{2}}
(:e^{(a+b)\phi(u)}:+\sum_{k=1}^{\infty}d_k(u)(iu-iu')^k),\nonumber
\end{eqnarray}
where $c_k(u)$ and $d_k(u)$ are operator-valued functions of $u$.
Now one can see that only two types of terms 
can give the contribution of the order $\epsilon$ in 
$\mathbf{\bar{L}}^{(q)}(x_{m-1},x_{m})$ when $q\to 1$.
The first type consists of operators of the first order in $V_i$ 
and the second type is formed
by the operators, quadratic in $V_i$, which give contribution of the order
$\epsilon^{1\pm\beta^2}$ by virtue of operator product expansion.
Let's look on the terms of the second type in detail.
At first we consider the terms 
appearing from the $R_0$-part of R-matrix, represented by vertex operators:
\begin{eqnarray}
&&\frac{e_{\delta}}{2(q-q^{-1})}\Bigg(\int^{x_m}_{x_{m-1}}\d u_1 :e^{-\phi}:
\xi(u_1-i0)\int^{x_m}_{x_{m-1}}\d u_2 :e^{\phi}:\xi(u_2+i0)+\nonumber\\
&&q^{-1}\int^{x_m}_{x_{m-1}}\d u_2 :e^{\phi}:
\xi(u_2-i0)\int^{x_m}_{x_{m-1}}\d u_1 :e^{-\phi}:\xi(u_1+i0)\Bigg)
\end{eqnarray}
Neglecting the terms, which give rise to $O(\epsilon^2)$ contribution, 
we obtain, using the operator products of vertex operators:
\begin{eqnarray}
&&\frac{e_{\delta}}{2(q-q^{-1})}\Bigg(\int^{x_m}_{x_{m-1}}\d u_1 
\int^{x_m}_{x_{m-1}}\d u_2 \frac{-i\beta^2}{(i(u_1-u_2-i0))^{\frac{\beta^2}{2}
+1}}+\\
&&q^{-1}\int^{x_m}_{x_{m-1}}\d u_2 \int^{x_m}_{x_{m-1}}\d u_1 
\frac{-i\beta^2}{(i(u_2-u_1-i0))^{\frac{\beta^2}{2}+1}}\Bigg)\nonumber
\end{eqnarray}
In the $\beta^2\to 0$ limit we get:
$
\frac{[e_{\alpha_1},e_{\alpha_0}]}{2i\pi}
\int^{x_m}_{x_{m-1}}\d u_1 
\int^{x_m}_{x_{m-1}}\d u_2 \big(\frac{1}{u_1-u_2+i0}
-\frac{1}{u_1-u_2-i0}\big)\nonumber
$
which, by the well known formula: $\frac{1}{x+i0}=P\frac{1}{x} -i\pi\delta(x)$
gives: $-\int^{x_m}_{x_{m-1}}\d u [e_{\alpha_1}, e_{\alpha_0}]$.
Another terms arise from the $R_{+}$ and $R_{-}$ parts of R-matrix 
and are very similar to each other:
\begin{eqnarray}
\frac{e_{\alpha_0}^2}{2(2)_{(-q^{-1})}}\int^{x_m}_{x_{m-1}}
\d u_1 :e^{\phi}:\xi(u_1-i0)
\int^{x_m}_{x_{m-1}}\d u_2 :e^{\phi}:\xi(u_2+i0),\\
\frac{e_{\alpha_1}^2}{2(2)_{(-q^{-1})}}\int^{x_m}_{x_{m-1}}
\d u_1 :e^{-\phi}:\xi(u_1-i0)
\int^{x_m}_{x_{m-1}}\d u_2 :e^{-\phi}:\xi(u_2+i0).\nonumber
\end{eqnarray}
The integrals can be reduced to the ordered ones: 
\begin{eqnarray}
\frac{e_{\alpha_0}^2}{2}\int^{x_m}_{x_{m-1}}\d u_1 :e^{\phi}:\xi(u_1)
\int^{u_1}_{x_{m-1}}\d u_2 :e^{\phi}:\xi(u_2),\\
\frac{e_{\alpha_1}^2}{2}\int^{x_m}_{x_{m-1}}\d u_1 :e^{-\phi}:\xi(u_1)
\int^{u_1}_{x_{m-1}}\d u_2 :e^{-\phi}:\xi(u_2).\nonumber
\end{eqnarray}
Following \cite{super-kdv} we find that their contribution (of order $\epsilon$) in 
the classical limit is:
\begin{eqnarray}
-e_{\alpha_0}^2\int^{x_m}_{x_{m-1}}\d u e^{2\phi(u)},\quad
-e_{\alpha_1}^2\int^{x_m}_{x_{m-1}}\d u e^{-2\phi(u)}.
\end{eqnarray} 
Gathering now all the terms of order $\epsilon$ we find:
\begin{eqnarray}
&&\mathbf{\bar{L}}^{(1)}(x_{m-1},x_{m})=1+\int^{x_m}_{x_{m-1}}\d u 
(\frac{i}{\sqrt{2}}
\xi(u)e^{-\phi(u)}e_{\alpha_1}-\\
&&\frac{i}{\sqrt{2}}
\xi(u)e^{\phi(u)}e_{\alpha_0}
-e^2_{\alpha_1}e^{-2\phi(u)}-
e^2_{\alpha_0}e^{2\phi(u)}-[e_{\alpha_1},e_{\alpha_0}])+O(\epsilon^2)
\nonumber
\end{eqnarray}
and collecting all $\mathbf{\bar{L}}^{(1)}(x_{m-1},x_{m})$ we find
that $\mathbf{\bar{L}}^{(1)}$=$e^{-i\pi p h_{\alpha_1}}\mathbf{L}$. Therefore 
$\mathbf{L}^{(1)}$=$\mathbf{L}$.

\section{$N=2$ SUSY KdV hierarchy in nonstandard form}
The matrix L-operator of the N=2 SUSY KdV has the following explicit form \cite{delduc}:
\begin{displaymath}
\mathcal{L}_F=D-
\left(\begin{array}{ccc}
D\Phi_1 & 1 & 0  \\
\lambda & D\Phi_1-D\Phi_2 & 1\\
\lambda D(\Phi_1- \Phi_2)& \lambda & -D\Phi_2\\
\end{array}\right),
\end{displaymath}
where $D=\partial_{\theta}+\theta\partial_u$ and $u$ is a variable on a cylinder of circumference 
$2\pi$ and $\Phi_i(u,\theta)=\phi_i(u)-\frac{i}{\sqrt{2}}\theta\xi_i(u)$ are the superfields with the following 
Poisson brackets:
\begin{eqnarray}
&&\{\Phi_i(u_1,\theta_1),\Phi_i(u_2,\theta_2)\}=0\quad (i=1,2)\nonumber\\
&&\{D_{u_1,\theta_1}\Phi_1(u_1,\theta_1),D_{u_2,\theta_2}
\Phi_2(u_2,\theta_2)\}=-D_{u_1,\theta_1}(\delta(u_1-u_2)(\theta_1-\theta_2))
\end{eqnarray} 
This is odd L-operator, related to the $sl^{(1)}(2|1)$ affine superalgebra (the elements in the associated column 
of the representation vector are graded from upper to the lower in such an order: even, odd, even). 
There is also one (canonical) related to the $sl^{(1)}(2|2)$ \cite{inami}. The quantization of 
this higher rank case can be performed by means of the procedure outlined in \cite{toda-kdv} and does not add 
something new in the quantization procedure with respect to \cite{super-kdv}, \cite{toda-kdv}. Moreover if we look forward to study and possibly solve 
the corresponding integrable quantum model we will find that the associated 
representation theory of $sl^{(1)}(2|2)$ (both classical and quantum) is very complicated. 
However the introduced lower rank form allows us to work with much more simple representation theory 
of $sl^{(1)}(2|1)$ and, as we will see, it provides very interesting features of the 
quantization formalism.\\     
\hspace*{5mm}
In order to build the scalar L-operator related with this matrix one, 
let's consider the corresponding 
linear problem ($\mathcal{L}\Psi=0$) and express the second an third element in the 
vector $\Psi$ in terms of the first (upper) one. 
The linear equation for this element is the following one:
\begin{eqnarray} 
((D+D\Phi_1)(D-D(\Phi_1-\Phi_2))(D-D\Phi_2)+2\lambda D)\Psi_1=0
\end{eqnarray}
Therefore the scalar linear operator is:
\begin{eqnarray}\label{fermil} 
L=D^3+(\mathcal{V}+2\lambda)D+\mathcal{U},
\end{eqnarray}
where the Miura map is given by:
\begin{eqnarray}\label{pot}
&&\mathcal{V}=-D\Phi_2 D\Phi_1+\partial\Phi_2-\partial\Phi_1=
\frac{i}{\sqrt{2}}\theta(\alpha^+-\alpha^-)  +V,\nonumber\\
&&\mathcal{U}=-\partial\Phi_2 D\Phi_1-D\partial\Phi_1=\theta U+\frac{i}{\sqrt{2}}\alpha^+\nonumber\\
&&U=-\phi_1'\phi_2'-\frac{1}{2}\xi_1\xi_2'-\phi_1'', \quad \alpha^+=\xi_1\phi_2'+\xi_1',\nonumber\\ 
&&\alpha^-=\xi_2\phi_1'+\xi_2', \quad V=\phi_2'-\phi_1'+\frac{1}{2}\xi_2\xi_1
\end{eqnarray}  
and the introduced objects $U,V,\alpha^{\pm}$ satisfy the N=2 superconformal algebra under the 
Poisson brackets:
\begin{eqnarray}\label{poisson} 
&&\{U(u),\alpha^+(v)\}=-\alpha'^+(u)\delta(u-v)-2\alpha^+(u)\delta'(u-v),\nonumber\\
&&\{U(u),\alpha^-(v)\}=-\alpha^-(u)\delta'(u-v),\nonumber\\
&&\{\alpha^+(u),\alpha^-(v)\}=-2U(u)\delta(u-v)-2V(u)\delta'(u-v)-2\delta''(u-v),\nonumber\\
&&\{V(u), \alpha^+(v)\}=-\alpha^+(u) \delta(u-v),\nonumber\\
&&\{V(u), \alpha^-(v)\}=\alpha^-(u) \delta(u-v),\nonumber\\
&&\{U(u),V(v)\}=-V(u)\delta'(u-v),\nonumber\\
&&\{V(u),V(v)\}=-2\delta'(u-v),\nonumber\\
&&\{U(u),U(v)\}=-U'(u)\delta(u-v)-2U(u)\delta'(u-v).
\end{eqnarray}  
Now let's rewrite the above L-operator in the purely algebraic form. That is: 
\begin{eqnarray}
&&\mathcal{L}_F=D-(h_{\alpha_1}D\Phi_1+h_{\alpha_2}D\Phi_2+e_{\alpha_1}+e_{\alpha_2}+[e_{\alpha_2},
e_{\alpha_0}]+ \nonumber\\
&&[e_{\alpha_0}, e_{\alpha_1}]+D(\Phi_1- \Phi_2)e_{\alpha_0})
\end{eqnarray}
where $h_{\alpha_i},e_{\alpha_i}$ are the generators of the upper Borel algebra of  
$sl^{(1)}(2|1)$, with the following commutation relations:  
\begin{eqnarray}
&&[e_{\alpha_i}, e_{-\alpha_j}]=\delta_{i,j}h_{\alpha_i}\quad (i=0,1,2), \quad
[h_{\alpha_j},e_{\pm\alpha_i}]=\pm e_{\pm\alpha_i} \quad (i=1,2, i \neq j)\nonumber\\
&&[h_{\alpha_j},e_{\pm\alpha_0}]=\mp e_{\pm\alpha_0} \quad (i=1,2),\quad
[h_{\alpha_0},e_{\pm\alpha_i}]=\mp e_{\pm\alpha_i} \quad (i=1,2)\nonumber\\
&&[h_{\alpha_i},e_{\pm\alpha_i}]=0 \quad (i=1,2),\quad
[h_{\alpha_0},e_{\pm\alpha_0}]=\pm 2e_{\pm\alpha_0},\nonumber\\
&&ad^2_{e_{\pm\alpha_i}}e_{\pm\alpha_j}=0,\quad [e_{\pm\alpha_k},e_{\pm\alpha_k}] =0 \quad (k=1,2)
\end{eqnarray}
where $e_{\pm\alpha_i}$ are odd generators for $i=1,2$ and even for $i=0$; $[,]$ means the 
supercommutator.\\ 
The symmetrized Cartan's matrix $b_{ij}=(\alpha_i,\alpha_j)$ corresponding to the given affine superalgebra is equal to:
\begin{eqnarray}
b=
\left(\begin{array}{ccc}
0 & 1 & -1  \\
1 & 0 & -1\\
-1& -1 & 2\\
\end{array}\right),
\end{eqnarray}
Throughout the paper we will use only the evaluation representations (this means that central 
charge of the corresponding affine algebra in these representations is equal to zero). 
In order to write a monodromy matrix we need to consider the equivalent bosonic L-operator. That is, 
expressing the linear problem associated with the operator (\ref{fermil}) in the following 
way: $\mathcal{L}\Psi=(\partial_{\theta}+\theta\partial_u+N_1+\theta N_0)(\Psi_0+\theta\Psi_1)$, 
it is easy to rewrite it as a bosonic linear problem on $\Psi_0$: $\mathcal{L}_B\Psi_0\equiv
(\partial_u+N^2_1+N_0)\Psi_0=0$ and $\Psi_1=-N_1 \Psi_0$.
where $N_1=\frac{i}{\sqrt{2}}\xi_1 h_{\alpha_1}+\frac{i}{\sqrt{2}}\xi_2 h_{\alpha_2}- 
e_{\alpha_1}-e_{\alpha_2}-[e_{\alpha_2},e_{\alpha_0}] - [e_{\alpha_0}, e_{\alpha_1}]+
\frac{i}{\sqrt{2}}(\xi_1-\xi_2)e_{\alpha_0}$ and 
$N_0=-\phi_1'h_{\alpha_1} - \phi_2'h_{\alpha_2}-(\phi_1'-\phi_2')e_{\alpha_0}$. that is:
\begin{eqnarray}
&&\mathcal{L}_B=\partial_u + (\frac{i}{\sqrt{2}}\xi_1h_{\alpha_1}+
\frac{i}{\sqrt{2}}\xi_2h_{\alpha_2}-[e_{\alpha_2},e_{\alpha_0}]-
[e_{\alpha_0},e_{\alpha_1}]\nonumber\\
&&-e_{\alpha_1}-e_{\alpha_2}+
\frac{i}{\sqrt{2}}(\xi_1-\xi_2)e_{\alpha_0})^2-(\phi'_1-\phi'_2)e_{\alpha_0}-
\phi'_1h_{\alpha_1}-\phi'_2h_{\alpha_2}
\end{eqnarray}
Considering the associated linear problem one can write the solution in the following way:
\begin{eqnarray}
\chi(u)=e^{\sum_{i=1,2}\phi_i(u)h_{\alpha_i}}Pexp\int_0^{u}\d u'(\sum_{k=0,1,2}
W_{\alpha_k}(u')e_{\alpha_k}+K(u'))
\end{eqnarray}  
where $W_{\alpha_i}=\int \d\theta e^{-\Phi_i}$ for $i=1,2$ and 
$W_{\alpha_0}=\int \d\theta (D\Phi_1-D\Phi_2)e^{\Phi_1+\Phi_2}$,
\begin{eqnarray}\label{K}
&&K(u)=-\frac{i}{\sqrt{2}}\xi_2[e_{\alpha_2},e_{\alpha_0}]e^{\phi_2}-
\frac{i}{\sqrt{2}}\xi_1[e_{\alpha_0},e_{\alpha_1}]e^{\phi_1}-[e_{\alpha_1},e_{\alpha_2}]e^{-\phi_1-\phi_2}
-\nonumber\\
&&[[e_{\alpha_0},e_{\alpha_1}],[e_{\alpha_2},e_{\alpha_0}]]e^{\phi_1+\phi_2}-
[e_{\alpha_2},[e_{\alpha_0},e_{\alpha_1}]]-
[e_{\alpha_1},[e_{\alpha_2},e_{\alpha_0}]].
\end{eqnarray}
We can define then the monodromy matrix on the interval $[0,2\pi]$ with quasiperiodic boundary conditions
($\phi_i(u+2\pi)=\phi_i(u)+2\pi ip_i$, $\xi_i(u+2\pi)=\pm\xi_i(u)$) in the following way:
\begin{eqnarray}\label{moncl}
\mathbf{M}=e^{\sum_{i=1,2}2i\pi p_ih_{\alpha_i}}Pexp\int_0^{2\pi}\d u(\sum_{k=0,1,2}
W_{\alpha_k}(u)e_{\alpha_k}+K(u))
\end{eqnarray}  
The reason of separating the terms in the $K$-part and the covariant exponential superfield part is that in the 
quantum case (see below) the complicated and 
noncovariant $K$-part disappears from the expression of the quantum monodromy matrix. 
As in the case of the standard KdV models let's define the auxiliary L-operators
\begin{eqnarray}\label{L}
\mathbf{L}=e^{-\sum_{i=1,2}i\pi p_ih_{\alpha_i}}\mathbf{M},
\end{eqnarray}  
which satisfy the quadratic r-matrix relation \cite{solitons}:  
\begin{eqnarray}
\{\mathbf{L}(\lambda),\otimes \mathbf{L}(\mu)\}=[\mathbf{r}(\lambda\mu^{-1}), 
\mathbf{L}(\lambda)\otimes\mathbf{L}(\mu)]
\end{eqnarray}
where we have restored the dependence on the spectral parameters 
$\lambda,\mu$ corresponding to some evaluation representations. Here $\mathbf{r}(\lambda\mu^{-1})$ 
is a classical trigonometric r-matrix associated with $sl^{(1)}(2|1)$ 
\cite{shadr}.
As usual this leads to the classical integrability relation:
\begin{eqnarray}
\{\mathbf{t}(\lambda),\mathbf{t}(\mu)\}=0
\end{eqnarray}     
where $\mathbf{t}(\lambda)=str\mathbf{M}(\lambda)$ and the supertrace is 
taken in some evaluation representation of $sl^{(1)}(2|1)$.

\section{Quantum R-matrix and the Cartan-Weyl basis for $sl_q^{(1)}(2|1)$} 
The quantum algebra $sl_q^{(1)}(2|1)$ has the following commutation relations:
\begin{eqnarray}
&&[e_{\alpha_i}, e_{-\alpha_j}]=\delta_{i,j}[h_{\alpha_i}]\quad (i=0,1,2), \quad
[h_{\alpha_j},e_{\pm\alpha_i}]=\pm e_{\pm\alpha_i} \quad (i=1,2, i \neq j)\nonumber\\
&&[h_{\alpha_j},e_{\pm\alpha_0}]=\mp e_{\pm\alpha_0} \quad (i=1,2),\quad
[h_{\alpha_0},e_{\pm\alpha_i}]=\mp e_{\pm\alpha_i} \quad (i=1,2)\nonumber\\
&&[h_{\alpha_i},e_{\pm\alpha_i}]=0 \quad (i=1,2),\quad
[h_{\alpha_0},e_{\pm\alpha_0}]=\pm 2e_{\pm\alpha_0},\nonumber\\
&&\label{serre}[e_{\pm\alpha_i},[e_{\pm\alpha_i},e_{\pm\alpha_j}]_q]_q=0,\quad e^2_{\alpha_k}=0 \quad (k=1,2).
\end{eqnarray} 
where the quantum supercommutator is $[e_{\alpha},e_{\beta}]_q=
e_{\alpha}e_{\beta}-(-1)^{p(\alpha)p(\beta)}q^{(\alpha,\beta)}e_{\beta}e_{\alpha}$ and 
$[x]=\frac{q^x-q^{-x}}{q-q^{-1}}$.\\ 
\hspace*{5mm}The expression for the quantum R-matrix for quantum affine superalgebras is 
\cite{khor}:
\begin{equation}\label{R-matrix}
R=K\bar{R}=K(\prod^{\to}_{\alpha\in{\Delta_{+}}}R_{\alpha}),
\end{equation}
where $\bar{R}$ is a reduced R-matrix and $R_{\alpha}$ are defined by the formulae:
\begin{equation}
R_{\alpha}=exp_{q_{\alpha}^{-1}}((-1)^{p(\alpha)}(q-q^{-1})(a(\alpha))^{-1}(e_{\alpha}\otimes e_{-\alpha}))
\end{equation}
for real roots and 
\begin{equation}
R_{n\delta}=exp((q-q^{-1})(\sum^{mult}_{i,j}c_{ij}(n)
e^{(i)}_{n\delta}\otimes e^{(j)}_{-n\delta}))
\end{equation}
for pure imaginary roots. 
Here $\Delta_{+}$ is the reduced positive root system (the bosonic roots which are two 
times fermionic roots are excluded) and q-exponen\-tial is defined as usual: 
$$ exp_q(x)=
\sum^{\infty}_{n=0} x^n/(n)_q!,$$ 
where 
$$(n)_q=(q^n-1)/(q-1).$$ 
The generators corresponding to the composite roots and the ordering in 
(\ref{R-matrix}) are defined according to the construction of the Cartan-Weyl basis (see e.g. \cite{khor}). 
The Cartan factor $K$ in the case of the $sl_q^{(1)}(1|2)$ is equal to :
$$
K=q^{h_{\alpha_1}\otimes h_{\alpha_2}+h_{\alpha_2}\otimes h_{\alpha_1}}
$$ 
The $a(\alpha), c_{ij}(n), d_{ij}(n)$ coefficients and are defined as follows:
$$
[e_{\gamma},e_{-\gamma}]=a(\gamma)(k_{\gamma}-k^{-1}_{\gamma})(q-q^{-1}), 
$$ $$[e^{(i)}_{n\delta},e^{(j)}_{-n\delta}]=d_{ij}(n)(q^{nh_{\delta}}-q^{-nh_{\delta}})/(q-q^{-1})$$ and 
$c_{ij}(n)$ is an inverse matrix to $d_{ij}(n)$.
The first few generators corresponding to composite roots and constructed by means of the 
mentioned procedure are:
\begin{eqnarray}
&&e_{\alpha_1+\alpha_2}=[e_{\alpha_1},e_{\alpha_2}]_{q^{-1}}\\
&&e_{\delta-\alpha_1}\equiv e_{\alpha_0+\alpha_2}= [e_{\alpha_0},e_ {\alpha_2}]_{q^{-1}},\quad 
e_{\delta-\alpha_2}\equiv e_{\alpha_0+\alpha_1}= [e_{\alpha_1},e_ {\alpha_0}]_{q^{-1}}\nonumber\\
&&e^{(1)}_{\delta}=[[e_{\alpha_0},e_ {\alpha_2}]_{q^{-1}},e_{\alpha_1}],\quad 
e^{(2)}_{\delta}=[[e_{\alpha_1},e_ {\alpha_0}]_{q^{-1}},e_ {\alpha_2}]\nonumber\\
&&e_{2\delta-\alpha_1-\alpha_2}=[e_{\delta-\alpha_2},e_{\delta-\alpha_1}]_{q^{-1}}\nonumber\\
&&e_{-\alpha_1-\alpha_2}=[e_{-\alpha_2},e_{-\alpha_1}]_{q}\nonumber\\
&&e_{-\delta+\alpha_1}\equiv e_{-\alpha_0-\alpha_2}= [e_{-\alpha_2},e_ {-\alpha_0}]_{q},\quad 
e_{-\delta+\alpha_2}\equiv e_{-\alpha_0-\alpha_1}= [e_{-\alpha_0},e_ {-\alpha_1}]_{q}\nonumber\\
&&e^{(1)}_{-\delta}=[[e_{-\alpha_2},e_ {-\alpha_0}]_{q},e_{-\alpha_1}],\quad 
e^{(2)}_{-\delta}=[[e_{-\alpha_0},e_ {-\alpha_1}]_{q},e_{-\alpha_2}]\nonumber\\
&&e_{-2\delta+\alpha_1+\alpha_2}=[e_{-\delta+\alpha_1},e_{-\delta+\alpha_2}]_{q}\nonumber
\end{eqnarray}

\section{The Construction of the Quantum Monodromy Matrix}
In this section the quantum version of the monodromy matrix introduced in Sec. 2 will be constructed. 
It will be shown that in classical limit one obtains the right answer, that is the expression 
(\ref{moncl}).\\
\hspace*{5mm}At first let's consider the quantum versions of the integrated exponentials of the fields 
(vertex operators):
\begin{eqnarray}\label{vertex}
&&W_{\alpha_i}=\int \d\theta :e^{-\Phi_i}:\equiv \frac{i}{\sqrt{2}}\xi_i :e^{-\phi_i}: \quad(i=1,2),\nonumber\\
&&W_{\alpha_0}=\int \d\theta :(D\Phi_1-D\Phi_2)e^{\Phi_1+\Phi_2}:\equiv :e^{\phi_1+\phi_2}(\phi'_1-
\phi'_2+\xi_1\xi_2):.
\end{eqnarray}
One can express $\Phi_1=\frac{i\Phi_{+}+\Phi_{-}}{\sqrt{2}}, \quad 
\Phi_2=\frac{i\Phi_{+}-\Phi_{-}}
{\sqrt{2}}$, where as usual:
\begin{eqnarray}
&&\phi_{\pm}(u)=iQ^{\pm}+iP^{\pm}u+\sum_n\frac{a^{\pm}_{-n}}{n}e^{inu},\qquad
\xi_{\pm}(u)=i^{-1/2}\sum_n\xi^{\pm}_ne^{-inu},\nonumber\\
&&[Q^{\pm},P^{\pm}]=\frac{i}{2}\beta^2 ,\quad 
[a^{\pm}_n,a^{\pm}_m]=\frac{\beta^2}{2}n\delta_{n+m,0},\qquad
\{\xi^{\pm}_n,\xi^{\pm}_m\}=\beta^2\delta_{n+m,0}.\nonumber
\end{eqnarray}
and the normal ordering in (\ref{vertex}) is defined as usual:\\
$
:e^{c\phi_{\pm}(u)}:=
\exp\Big(c\sum_{n=1}^{\infty}\frac{a^{\pm}_{-n}}{n}e^{inu}\Big)
\exp\Big(ci(Q^{\pm}+P^{\pm}u)\Big)\exp\Big(-c\sum_{n=1}^{\infty}\frac{a^{\pm}_{n}}{n}e^{-inu}.
\Big)
$
Here the $a^{\pm}_{n}$ operators with $n$ negative are placed to the left and with $n$ positive to 
the right.   
The vertex operators (\ref{vertex}) integrated from $u_1$ to $u_2$ satisfy the quantum Serre and 
``non Serre'' relations (\ref{serre}) for the lower Borel subalgebra with 
$q=e^{\frac{i\pi\beta^2}{2}}$ \cite{feigin}. 

The proof of this proposition is nontrivial because the usual proof of the Serre relations 
given in the papers \cite{feigin} for the bosonic case, based on the transformation of 
the product of the integrated vertex operators to the ordered integrals is not valid here. 
This happens because of singularities generated by the fermion fields in the corresponding 
operator products. But there is another way of proof, which uses usual in CFT technique of 
contour integration and analytic continuation of the operator products of nonlocal vertex 
operators \cite{klevtsov}. 
This proof is appropriate also for the case of quantum affine superalgebra and corresponding 
vertex-operators, because this method allows to pick out the divergences and then to cancel 
them in the Serre and ``nonstandard'' Serre relations. This problem will be considered in 
the separate paper.     

This allows us to represent 
the $e_{-\alpha_i}$ 
generators by the $(q-q^{-1})^{-1}\int^{u_2}_{u_1}W_{\alpha_i}$ operators. It was shown in \cite{bhk} that the 
corresponding reduced R-matrix $\bar{R}$, which we will denote 
$\bar{\mathbf{L}}^{(q)}(u_2,u_1)$ enjoys 
the P-exponential property, satisfying the following functional relation:  
\begin{eqnarray}
\bar{\mathbf{L}}^{(q)}(u_3,u_1)=\bar{\mathbf{L}}^{(q)}(u_3,u_2)
\bar{\mathbf{L}}^{(q)}(u_2,u_1).
\end{eqnarray}
However, in the supersymetric case when the fermion operators appear the associated 
singularities in the operator products does not allow us to write it as usual, in terms of 
ordered integrals. 
That's why we have called it ``quantum'' P-exponential, which in our case can be written in 
the following way:
\begin{eqnarray}
\bar{\mathbf{L}}^{(q)}(u_2,u_1)=Pexp^{(q)}\int_{u_1}^{u_2}\d u(\sum_{k=0,1,2}
W_{\alpha_k}(u)e_{\alpha_k})
\end{eqnarray}
One can show that $e^{\sum_{i=1,2}\pi i p_ih_{\alpha_i}}\bar{\mathbf{L}}^{(q)}(2\pi,0)=
\mathbf{L}^{(q)}$ satisfy the RTT-relation \cite{leshouches}, \cite{kulsklyan}:
\begin{eqnarray}
&&\mathbf{R}(\lambda\mu^{-1})
\Big(\mathbf{L}^{(q)}(\lambda)\otimes \mathbf{I}\Big)\Big(\mathbf{I}
\otimes \mathbf{L}^{(q)}(\mu)\Big)\\
&&=(\mathbf{I}\otimes \mathbf{L}^{(q)}(\mu)\Big)
\Big(\mathbf{L}^{(q)}(\lambda)\otimes \mathbf{I}\Big)\mathbf{R}
(\lambda\mu^{-1}), \nonumber
\end{eqnarray}
where the dependence on $\lambda,\mu$ means that we are considering
$\mathbf{L}^{(q)}$-operators in the evaluation representation of $sl_q^{(1)}(2|1)$.
We note also, that $\mathbf{M}^{(q)}=e^{\sum_{i=1,2}\pi i p_ih_{\alpha_i}}\bar{\mathbf{L}}^{(q)}$ satisfies the reflection equation \cite{reflection}:
\begin{eqnarray}
\mathbf{\tilde{R}}_{12}(\lambda\mu^{-1})
\mathbf{M}_1^{(q)}(\lambda)F^{-1}_{12}\mathbf{M}^{(q)}_2(\mu)=\mathbf{M}_2^{(q)}(\mu)F^{-1}_{12}
\mathbf{M}_1^{(q)}(\lambda)\mathbf{R}_{12}(\lambda\mu^{-1}),
\end{eqnarray}
where $F=K^{-1}$, the Cartan's factor from the universal R-matrix, 
$\mathbf{\tilde{R}}_{12}(\lambda\mu^{-1})=F^{-1}_{12}\mathbf{R}_{12}(\lambda\mu^{-1})F_{12}$ and labels 1,2 denote the position of multipliers in the tensor 
product. This leads to the quantum integrability relation:
\begin{eqnarray}
[\mathbf{t}^{(q)}(\lambda), \mathbf{t}^{(q)}(\mu)]=0,
\end{eqnarray} 
where $\mathbf{t}^{(q)}(\lambda)=str\mathbf{M}^{(q)}(\lambda)$.\\
\hspace*{5mm}Now we want to show that in the classical limit when $q\to 1$ $\mathbf{L}^{(q)}$, $\mathbf{M}^{(q)}$ 
give the auxiliary $\mathbf{L}$-matrix and the monodromy matrix 
correspondingly. 
Note also, that as usual, quantum universal R-matrix gives the classical r-matrix in the limit 
$q\to 1$, therefore the classical limit of  $\mathbf{L}^{(q)}$-operator gives the realization 
of classical r-matrix by means of classical counterparts of the corresponding vertex operators.

In order to find the classical limit, 
let's decompose $\mathbf{\bar{L}}^{(q)}$ in the following way 
\cite{toda-kdv}:
\begin{equation}
\mathbf{\bar{L}}^{(q)}(2\pi,0)=
\lim_{N\to\infty}\prod_{m=1}^{N}\mathbf{\bar{L}}^{(q)}(x_{m},x_{m-1}), 
\end{equation}
where we divided  the interval $[0,2\pi]$ into infinitesimal 
intervals $[x_m,x_{m+1}]$
with $x_{m+1}-x_m=\epsilon=2\pi/N$.
Let's find the terms that can give contribution of the first order
in $\epsilon$ in $\mathbf{\bar{L}}^{(q)}(x_{m},x_{m-1})$. 
In this analysis we will need the operator products of 
vertex operators. The ones giving nontrivial terms are:
\begin{eqnarray}
&&\xi_1(u)\xi_2(u')=
\frac{i\beta^2}{(iu-iu')}+\sum_{k=0}^{\infty}c_k(u)(iu-iu')^k,\nonumber\\
&&:e^{a\phi_1(u)}::e^{b\phi_2(u')}:=(iu-iu')^{\frac{-ab\beta^2}{2}}
(:e^{(a\phi_1(u)+b\phi_2(u)}:+\nonumber\\
&&\sum_{k=1}^{\infty}d_k(u)(iu-iu')^k),\nonumber\\
&&\phi_1'(u):e^{b\phi_2(u')}:=\frac{-ib\beta^2:e^{b\phi_2(u)}:}{2(iu-iu')}+
\sum_{k=0}^{\infty}f_k(u)(iu-iu')^k,\nonumber\\
&&\phi_2'(u):e^{b\phi_1(u')}:=\frac{-ib\beta^2:e^{b\phi_1(u)}:}{2(iu-iu')}+
\sum_{k=0}^{\infty}f_k(u)(iu-iu')^k.
\end{eqnarray}
In the previous cases considered in \cite{toda-kdv} only two types of terms 
gave the contribution of the order $\epsilon$ in 
$\mathbf{\bar{L}}^{(q)}(x_{m-1},x_{m})$ when $q\to 1$.
The first type consists of operators of the first order in $W_{\alpha_i}$ 
and the second type is formed
by the operators, quadratic in $W_{\alpha_i}$, which give contribution of the order
$\epsilon^{1\pm\beta^2}$ by virtue of operator product expansion. This second type contributions correspond
 to the composite roots which are equal to the sum of two simple roots.\\ 
\hspace*{5mm}In the present case we will show that there are contributions of the composite roots which are 
equal to the sum of three and even four simple roots providing the desired terms in the classical 
expression (\ref{K}-\ref{moncl}).
First, let's consider the quadratic terms corresponding to the negative roots $-\alpha_1-\alpha_2$, 
$-\delta+\alpha_2$, $-\delta+\alpha_1$. The commutation relations between vertex operators on a circle are:
\begin{eqnarray}
&&W_{\alpha_i}(u)W_{\alpha_j}(u')=-q^{-1}W_{\alpha_j}(u')W_{\alpha_i}(u)
\quad u>u' \quad (i,j=1,2\quad i\neq j)\nonumber\\
&&W_{\alpha_i}(u)W_{\alpha_0}(u')=q W_{\alpha_0}(u')W_{\alpha_i}(u) 
\quad u>u' \quad (i=1,2) \nonumber\\
&&W_{\alpha_0}(u)W_{\alpha_i}(u')=q W_{\alpha_i}(u')W_{\alpha_0}(u) 
\quad u>u' \quad (i=1,2) 
\end{eqnarray}
This allows us to write the generators corresponding to the negative composite roots  
$-\delta+\alpha_2$, $-\delta+\alpha_1$ in the following way:
\begin{eqnarray}
&&[e_{-\alpha_0}, e_{-\alpha_1}]_q=
\frac{1}{q-q^{-1}}\int_{x_{m-1}}^{x_{m}}\d uW_{\alpha_1}(u)\int_{x_{m-1}}^u 
\d u'W_{\alpha_0}(u')\nonumber\\
&&[e_{-\alpha_2}, e_{-\alpha_0}]_q=
\frac{1}{q-q^{-1}}\int_{x_{m-1}}^{x_{m}}\d uW_{\alpha_0}(u)\int_{x_{m-1}}^u \d u'W_{\alpha_2}(u')
\end{eqnarray}
The exponents of the corresponding q-exponentials from (\ref{R-matrix}) are equal to  
\begin{eqnarray}
&&\int_{x_{m-1}}^{x_{m}}\d uW_{\alpha_1}(u)\int_{x_{m-1}}^u \d u'W_{\alpha_0}(u')[e_{\alpha_1}, e_{\alpha_0}]_{q^{-1}}\nonumber\\
&&\int_{x_{m-1}}^{x_{m}}\d uW_{\alpha_0}(u)\int_{x_{m-1}}^u \d u'W_{\alpha_2}(u')[e_{\alpha_0}, e_{\alpha_2}]_{q^{-1}}
\end{eqnarray}
In the classical limit ($\beta^2\to 0$) their contribution due to the operator products (this calculation is 
similar to the calculation given in \cite{super-kdv}) is:
\begin{eqnarray}
\int_{x_{m-1}}^{x_{m}}\d u(-\frac{i}{\sqrt{2}}\xi_2e^{\phi_2}[e_{\alpha_2},e_{\alpha_0}]-
\frac{i}{\sqrt{2}}\xi_1e^{\phi_1}[e_{\alpha_0},e_{\alpha_1}])
\end{eqnarray}
The contribution of the root $-\alpha_1-\alpha_2$ is equal to   
\begin{eqnarray}
\int_{x_{m-1}}^{x_{m}}\d u(-e^{-\phi_1-\phi_2}[e_{\alpha_1},e_{\alpha_2}])
\end{eqnarray}
In this case the commutator of the integrated vertex operators cannot be rewritten in terms of the ordered 
integrals as it was above, however using the fact that the integrated vertex operators should be radially ordered 
(that is e.g. the product $e_{-\alpha_i}e_{-\alpha_j}$ should be written as $\int^{x_{m}}_{x_{m-1}}\d uW_{\alpha_i}(u-i0)
\int^{x_{m}}_{x_{m-1}}\d u'$ $W_{\alpha_i}(u'+i0)$)
 and the well known relation $\frac{1}{x+i0}-\frac{1}{x-i0}=-2i\pi\delta(x)$, we obtain this formula 
repeating calculation of \cite{toda-kdv}. Thus on this ``quadratic'' level we are in agreement with the classical 
expression.\\
\hspace*{5mm}Now let's look at the contribution corresponding to the composite roots $e^{(i)}_{\delta}$ and 
$2\delta-\alpha_1-\alpha_2$. 
First we consider the pure imaginary roots: 
$e^{(1)}_{-\delta}=[[e_{-\alpha_2},e_ {-\alpha_0}]_{q},e_{-\alpha_1}]$, 
$e^{(2)}_{-\delta}=[[e_{-\alpha_0},e_ {-\alpha_1}]_{q},e_ {-\alpha_2}]$. To calculate their contribution in the 
classical limit we at first need to take the terms which give  
contribution  of the order $\epsilon^{1+\beta^2}$ from 
$[e_{\alpha_2},e_ {\alpha_0}]_{q}$ and $[e_{\alpha_0},e_ {\alpha_1}]_{q}$, 
which are proportional to $\int \d u \d\theta e^{\Phi_2}$ and  $\int \d u \d\theta e^{\Phi_1}$. 
Then we consider their supercommutators with correspondingly $e_{\alpha_1}$ and $e_ {\alpha_2}$. 
Rewriting these supercommutators in terms of ordered integrals as it was above we find, following the 
calculations in \cite{super-kdv} that the contribution of the associated exponent of q-exponential is:
\begin{eqnarray}
\int^{x_{m}}_{x_{m-1}}\d u(-[e_{\alpha_2},[e_{\alpha_0},e_{\alpha_1}]]-
[e_{\alpha_1},[e_{\alpha_2},e_{\alpha_0}]])
\end{eqnarray}
Similarly, $e_{2\delta-\alpha_1-\alpha_2}$ generator is expressed as the q-commutator of 
$[e_{\alpha_2},e_ {\alpha_0}]_{q}$  and $[e_{\alpha_0},e_ {\alpha_1}]_{q}$, thus, leaving again their parts 
of order $\epsilon^{1+\beta^2}$ and using the formula $\frac{1}{x+i0}-\frac{1}{x-i0}=-2i\pi\delta(x)$ as 
it was explained above we obtain that the classical contribution of the first order in $\epsilon$ is:
\begin{eqnarray}
\int^{x_m}_{x_{m-1}}\d u(-[[e_{\alpha_0},e_{\alpha_1}],[e_{\alpha_2},e_{\alpha_0}]]e^{\phi_1+\phi_2})
\end{eqnarray}
Therefore, gathering all the terms we obtain that 
\begin{eqnarray}
&&\lim_{q\to 1}\mathbf{\bar{L}}^{(q)}(x_{m},x_{m-1})=1+\int^{x_m}_{x_{m-1}}\d u 
(\sum_{k=0,1,2}W_{\alpha_k}(u)e_{\alpha_k}-\frac{i}{\sqrt{2}}\xi_2[e_{\alpha_2},e_{\alpha_0}]e^{\phi_2}\nonumber\\
&&-
\frac{i}{\sqrt{2}}\xi_1[e_{\alpha_0},e_{\alpha_1}]e^{\phi_1}-[e_{\alpha_1},e_{\alpha_2}]e^{-\phi_1-\phi_2}
-[[e_{\alpha_0},e_{\alpha_1}],[e_{\alpha_2},e_{\alpha_0}]]e^{\phi_1+\phi_2}-\nonumber\\
&&[e_{\alpha_2},[e_{\alpha_0},e_{\alpha_1}]]-
[e_{\alpha_1},[e_{\alpha_2},e_{\alpha_0}]])
+O(\epsilon^2)
\end{eqnarray}
Collecting all the infinitesimal $\bar{\mathbf{L}}$ operators and multiplying on the appropriate Cartan factors 
we obtain the desired classical expressions for the $\mathbf{L}$-operators and the monodromy matrix.

\section{Conclusion}
\hspace*{5mm}
The construction of the monodromy matrix given in this paper follows the way introduced in 
\cite{1}, \cite{4}, \cite{susykdv}, \cite{toda-kdv}. 
It is shown that the quantum version of auxiliary $\mathbf{L}$-matrix coincides with the 
universal R-matrix with the lower Borel algebra represented by the appropriate 
vertex operators.
This construction allows also to demonstrate that the supersymmetry generators commute with the supertrace of the 
monodromy matrix, i.e. they can be included in the series of integrals of motion. 
Really, the commutators of the supersymmetry generator 
\begin{eqnarray}
G_0=\beta^{-2}\sqrt{2}i^{-1/2}\int_0^{2\pi}du \phi'(u) \xi(u)\nonumber
\end{eqnarray}
in the case of N=1 SUSY KdV and two supersymmetry generators 
\begin{eqnarray}
G^+_0=\beta^{-2}\sqrt{2}i^{-1/2}\int_0^{2\pi}du \phi_1'(u) \xi_2(u), \quad 
G^-_0=\beta^{-2}\sqrt{2}i^{-1/2}\int_0^{2\pi}du \phi_2'(u) \xi_1(u)\nonumber
\end{eqnarray}
in the case of N=2 SUSY KdV, with vertex operators from the corresponding quantum P-exponent 
reduce to the total derivatives and therefore 
(according to the results of \cite{toda-kdv},\cite{4}) they commute with the transfer matrices. 
The same way one can show that the transfer matrices commute with the zero mode of the U(1) current which 
is present 
in the $N=2$ superconformal algebra (the quantum version of the $V$ field from (\ref{pot})).\\  
\hspace*{5mm}This result has very important consequences. The SUSY N=1 KdV model can be used to 
describe the integrable structure of supersymmetric perturbed nonconformal two-dimensional integrable models 
with perturbation described by the vertex operator $\int\ud \theta e^{-\Phi} $ with the dimension $h_{1,3}$ 
from the Kac table.\\ 
\hspace*{5mm}In the case of N=2 SUSY KdV the situation is more interesting. 
If we make the twist transformation \cite{twist} in the 
underlying N=2 superconformal algebra we obtain one of the generators $G^{\pm}_0$ to become a BRST operator. 
It means that the transfer-matrices become BRST-exact providing the infinite series of pairwise commuting 
``physical'' integrals of 
motion (of the zero ghost number provided by the zero mode of U(1) current). 
This gives possibility to study the two-dimensional topological models and their integrable perturbations 
by means of methods of integrable field theories e.g. the most popular one, 
the so-called Quantum Inverse Scattering Method \cite{leshouches}, \cite{kulsklyan}.

\section*{Acknowledgments}
This work was supported by 
CRDF Grant RUM1-2622-ST-04 and RFBR Grant 05-01-00922. 
A.M.Z. is grateful to the Dynasty foundation for support and 
to N. Reshetikhin, M. Semenov-Tian-Shansky and 
A. Tseytlin for encouragement.

\end{document}